Video Article

# Bringing the Visible Universe into Focus with Robo-AO

Christoph Baranec[1,2], Reed Riddle[1], Nicholas M. Law[3], A.N. Ramaprakash[4], Shriharsh P. Tendulkar[2], Khanh Bui[1], Mahesh P. Burse[4], Pravin Chordia[4], Hillol K. Das[4], Jack T.C. Davis[1], Richard G. Dekany[1], Mansi M. Kasliwal[5], Shrinivas R. Kulkarni[1,2], Timothy D. Morton[2], Eran O. Ofek[6], Sujit Punnadi[4]

[1]Caltech Optical Observatories, California Institute of Technology
[2]Department of Astronomy, California Institute of Technology
[3]Dunlap Institute for Astronomy and Astrophysics, University of Toronto
[4]Inter-University Centre for Astronomy & Astrophysics
[5]Observatories of the Carnegie Institution for Science
[6]Benoziyo Center for Astrophysics, Weizmann Institute of Science

Correspondence to: Christoph Baranec at baranec@astro.caltech.edu

## Abstract

The angular resolution of ground-based optical telescopes is limited by the degrading effects of the turbulent atmosphere. In the absence of an atmosphere, the angular resolution of a typical telescope is limited only by diffraction, *i.e.*, the wavelength of interest, $\lambda$, divided by the size of its primary mirror's aperture, *D*. For example, the Hubble Space Telescope (HST), with a 2.4-m primary mirror, has an angular resolution at visible wavelengths of ~0.04 arc seconds. The atmosphere is composed of air at slightly different temperatures, and therefore different indices of refraction, constantly mixing. Light waves are bent as they pass through the inhomogeneous atmosphere. When a telescope on the ground focuses these light waves, instantaneous images appear fragmented, changing as a function of time. As a result, long-exposure images acquired using ground-based telescopes - even telescopes with four times the diameter of HST - appear blurry and have an angular resolution of roughly 0.5 to 1.5 arc seconds at best.

Astronomical adaptive-optics systems compensate for the effects of atmospheric turbulence. First, the shape of the incoming non-planar wave is determined using measurements of a nearby bright star by a wavefront sensor. Next, an element in the optical system, such as a deformable mirror, is commanded to correct the shape of the incoming light wave. Additional corrections are made at a rate sufficient to keep up with the dynamically changing atmosphere through which the telescope looks, ultimately producing diffraction-limited images.

The fidelity of the wavefront sensor measurement is based upon how well the incoming light is spatially and temporally sampled[1]. Finer sampling requires brighter reference objects. While the brightest stars can serve as reference objects for imaging targets from several to tens of arc seconds away in the best conditions, most interesting astronomical targets do not have sufficiently bright stars nearby. One solution is to focus a high-power laser beam in the direction of the astronomical target to create an artificial reference of known shape, also known as a 'laser guide star'. The Robo-AO laser adaptive optics system[2,3] employs a 10-W ultraviolet laser focused at a distance of 10 km to generate a laser guide star. Wavefront sensor measurements of the laser guide star drive the adaptive optics correction resulting in diffraction-limited images that have an angular resolution of ~0.1 arc seconds on a 1.5-m telescope.

## Video Link

The video component of this article can be found at http://www.jove.com/video/50021/

## Introduction

The impact of atmospheric turbulence on astronomical imaging was first recognized centuries ago by Christiaan Huygens[4] and Isaac Newton[5]. The first conceptual adaptive optics designs to compensate for the effects of turbulence were published independently by Horace Babcock[6] and Vladimir Linnik[7] in the 1950s. The U.S. Department of Defense then funded the development of the first adaptive-optics systems in the 1970s for the purpose of imaging foreign satellites during the Cold War[8]. The civilian astronomical community made progress developing systems in the 1980s; however, after declassification of military research on adaptive optics in 1992 (ref.[9]), there was an explosion in both the number and complexity of astronomical systems[10].

The majority of the approximately twenty visible and infrared telescopes today with apertures greater than 5 meters are equipped with adaptive-optics systems (*e.g.* refs. [11-19]). As telescopes become larger, and thus more capable at collecting light, there are greater gains in resolution and sensitivity when using adaptive optics. Unfortunately, large-telescope adaptive-optics systems are extremely complex and restricted in their operation to near-infrared wavelengths due to current technology; they require teams of support staff, often with large observing overheads, and access to these scarce and valuable resources is also limited.

At the other end of the size spectrum, there are well over one-hundred telescopes in the 1-3 meter class, but very few of these are fitted with adaptive optics. Correcting atmospheric turbulence, even at shorter visible wavelengths, becomes tractable with current technology on these smaller telescopes because they look through a much smaller volume of atmospheric turbulence (**Figure 1**). The total amount of turbulence-induced optical error scales almost proportionally with the telescope primary mirror diameter and inversely with the observing wavelength. The same adaptive-optics technology that is used with near-infrared light on the larger telescopes can be used with visible light on modest-sized telescopes. Additionally, many telescopes of this scale are either being retrofitted (*e.g.* ref. [20]) or newly built with fully robotic, remote and/or autonomous capabilities (*e.g.* ref. [21]), significantly increasing the cost-effectiveness of these facilities. If equipped with adaptive optics, these telescopes would offer a compelling platform to pursue many areas of astronomical science that are otherwise impractical or impossible with large telescope adaptive-optics systems[22]. Diffraction-limited targeted surveys of tens of thousands of targets[23,24], long-term monitoring[25,26], and rapid transient characterization in crowded fields[27], are possible with adaptive optics on these modest apertures.

To explore this new discovery space, we have engineered and implemented a new economical adaptive-optics system for 1-3 meter class telescopes, Robo-AO (refs. [2,3]; **Figure 2**). As with other laser adaptive-optics systems, Robo-AO comprises several main systems: the laser system; a set of electronics; and an instrument mounted at the telescope's Cassegrain focus (behind the primary mirror; **Figure 3**) that houses a high-speed optical shutter, wavefront sensor, wavefront correctors, science instruments and calibration sources. The Robo-AO design depicted herein illustrates how a typical laser adaptive-optics system operates in practice.

The core of the Robo-AO laser system is a Q-switched 10-W ultraviolet laser mounted in an enclosed projector assembly on the side of the telescope. Starting with the laser itself, the laser projector then incorporates a redundant shutter, in addition to the laser's internal shutter, for additional safety; a half-wave plate to adjust the angle of projected linear polarization; and an uplink tip-tilt mirror to both stabilize the apparent laser beam position on sky and to correct for telescope flexure. A bi-convex lens on an adjustable focus stage expands the laser beam to fill a 15 cm output aperture lens, which is optically conjugate to the tip-tilt mirror. The output lens focuses the laser light to a line-of-sight distance of 10 km. As the laser pulses (~35 ns long every 100 μs) propagate through the atmosphere away from the projector, a tiny fraction of the photons Rayleigh scatter off air molecules and return towards the telescope (**Figure 2B**). The returning scattered photons originate along the entire upward path of the laser, and would otherwise appear as a streak that would make the wavefront measurements inaccurate. Within the adaptive-optics instrument, a high-speed Pockels cell optical shutter[28] is used to transmit laser light only returning from just a narrow slice of the atmosphere around the 10 km projector focus, resulting in the laser appearing as a spot. Switching of the Pockels cell is driven by the same master clock as the pulsed laser, with a delay to account for the round trip time of the laser pulse through the atmosphere. Ultimately, only about one in every trillion photons launched is detected by the wavefront sensor. Even so, this radiant flux is sufficient to operate the adaptive-optics system.

The ultraviolet laser has the additional benefit of being invisible to the human eye, primarily due to absorption in the cornea and lens[29]. As such, it is unable to flash-blind pilots and is considered a Class 1 laser system (*i.e.* incapable of producing damaging radiation levels during operation and exempt from any control measures[30]) for all possible exposures of persons in overflying aircraft, eliminating the need for human spotters located on site as normally required by the Federal Aviation Authority within the U.S.[31]. Unfortunately, the possibility for the laser to damage some satellites in low Earth orbit may exist. For this reason, it is recommended for both safety and liability concerns to coordinate laser activities with an appropriate agency (*e.g.* with U.S. Strategic Command (USSTRATCOM) within the U.S.[32]).

The wavefront sensor which measures the incoming laser light within the Robo-AO Cassegrain instrument is known as a Shack-Hartmann sensor[33], and includes a lenslet array, optical relay and imaging sensor. The lenslet array is a refractive optical element, flat on one side, with a grid of square-shaped convex lenses on the other side. It is located at a position optically conjugate to the entrance pupil of the telescope. When the 'return light' from the laser passes through the lenslet array, images of the on-sky laser are created at the focus of each of the lenses in the array (**Figure 4**). This pattern of laser images is then optically relayed to a UV-optimized charge-coupled device (CCD) camera. The lateral x-y position of each image gives a measure of the local gradient or 'slope' of the light wave through each lens of the array. The signal-to-noise ratio of each position measurement with Robo-AO ranges from 6 to 10 depending on Zenith angle and seeing conditions (6.5 electrons of detector noise in each of four pixels with a signal ranging from 100 to 200 photoelectrons per image per measurement).

The overall shape of the light wave is then calculated by multiplying the measured slopes by a pre-computed wavefront reconstructor matrix. The reconstructor matrix is created by first making a model of the pupil geometry that is sub-divided by the lenslet array. Individual ortho-normal basis functions (in this case disk harmonic functions up to the $11^{th}$ radial order, for a total of 75 functions; ref. [34]) are realized over the model and a 2-D least-squares solution to the best-fit plane across each lens in the array is calculated. While this is an approximation to the average gradient, the difference is negligible in practice, with the benefit of easily handling the geometry of partially illuminated lenses at the edges of the projected pupil. An influence matrix is thus derived that converts unit amplitudes for each basis function with the slope offset for every lens. The reconstructor matrix is then created by taking the pseudo-inverse of the influence matrix using Singular Value Decomposition. Once the shape of the light wave is known in terms of coefficients of the basis set, a compensatory inverse shape can be commanded on the high-order wavefront corrector. The process of making a measurement, then applying a correction, and repeating this cycle over and over, is an example of an integral control-loop. Robo-AO executes its control-loop at a rate of 1.2 kHz, necessary to keep up with the dynamics of the atmosphere. A scale factor (also known as the gain of the integral control-loop) of less than 1, and typically close to 0.6, is applied to the correction signal to maintain the stability of the control-loop while still minimizing the residual error of corrected light.

The high-order wavefront corrector within Robo-AO is a micro-electro-mechanical-systems (MEMS) deformable mirror[35]. Robo-AO uses 120 actuators to adjust the illuminated surface of the mirror, sufficient in spatial resolution to accurately fit the calculated correcting shape. The actuators have a maximum surface deviation amplitude of 3.5 μm which corresponds to optical phase compensation of up to 7 μm. In typical atmospheric conditions at astronomical observatories, this compensation length is greater than 5 sigma of the amplitude of the turbulence induced optical error and therefore results in significant correction headroom. Furthermore, the deformable mirror can compensate for static optical errors arising from the instrument and telescope at the cost of reduced dynamic range.

One subtlety to using a laser as a probe of the atmosphere is its inability to measure astronomical image motion[36]. The returning laser light is viewed from roughly the same position from which it is projected and therefore should always appear in the same location on sky. Any overall tilt measured in the returning laser light wave by the wavefront sensor is dominated by mechanical pointing errors. The tilt signal is used to drive the laser system's uplink tip-tilt mirror, thus keeping the Shack-Hartmann pattern centered on the wavefront sensor. Correcting astronomical image motion is handled separately with the science cameras as explained below.

Robo-AO uses four off-axis parabolic (OAP) mirrors to relay light from the telescope to the science cameras achromatically (**Figure 3**). The relay path includes a fast tip-tilt correcting mirror as well as an atmospheric dispersion corrector (ADC)[37] comprised of a pair of rotating prisms. The ADC solves a particular issue related to observing objects through the atmosphere that are not directly overhead: the atmosphere acts as a prism and refracts light as a function of wavelength, with the overall effect becoming stronger as the telescope points lower in elevation, causing images - especially those that have been sharpened by adaptive optics correction - to appear elongated in the direction normal to the horizon. The ADC can add an opposite amount of dispersion to the incoming light, effectively negating the effect of the atmospheric prismatic dispersion (**Figure 5**). At the end of the OAP relay is a visible dichroic that reflects light of λ < 950 nm to an electron-multiplying charge-coupled device (EMCCD) camera while transmitting infrared light towards an infrared camera. The EMCCD camera has the ability to capture images with very low electronic (detector) noise[38,39], at a frame rate which reduces the intra-exposure image motion to below the diffraction-limited angular resolution. By re-centering and stacking a series of these images, a long-exposure image can be synthesized with minimal noise penalty. The EMCCD camera can also be used to stabilize image motion on the infrared camera; measurements of the position of an imaged astronomical source can be used to continuously command the fast tip-tilt to re-point the image to a desired location. Ahead of each camera is a set of filter wheels with an appropriate set of astronomical filters.

An internal telescope and source simulator is integrated into the Robo-AO system as a calibration tool. It can simultaneously simulate the ultraviolet laser focus at 10 km and a blackbody source at infinity, matching the host telescope's focal ratio and exit pupil position. The first fold mirror within Robo-AO directs all light from the telescope's secondary mirror to the adaptive-optics system. The fold mirror is also mounted on a motorized stage which can be translated out of the way to reveal the internal telescope and source simulator.

While the Robo-AO system is intended to operate in a completely autonomous fashion, each of the many steps of an adaptive optics observation can be executed manually. This step-by-step procedure, along with a brief explanation, is detailed in the following section.

## Protocol

### 1. Pre-observing Procedures

1. Make a list of the astronomical targets to be observed.
2. Calculate the total exposure times needed for each target to reach a required signal-to-noise-ratio in each scientific filter and camera combination desired.
3. Transmit the list of astronomical targets to be observed to USSTRATCOM greater than 3 days in advance of observations. They will send back a Predictive Avoidance Message (PAM) indicating 'open windows' - the times safe to use the laser system on each requested target without potentially damaging satellites.
4. Install the Robo-AO system on the telescope during the daytime if not already done (*e.g.* Robo-AO on the 1.5-m P60 telescope at Palomar Observatory, CA; **Figure 2**).
5. Translate the first fold mirror to reveal the internal telescope and source simulator to the laser wavefront sensor, and turn on the simulated laser source.
6. Record the positions of the simulated laser images on the wavefront sensor camera. These positions are used as reference slope measurements for the Shack-Hartmann wavefront sensor and will be subtracted from the following on-sky measurements. This procedure calibrates small optical changes in the instrument alignment due to changing temperatures.
7. Return the first fold mirror to its original position and turn off the simulated laser source.
8. Contact USSTRATCOM one hour prior to observing to inform them of the night's planned activity and receive any updates or changes to the PAM.
9. Turn the 10-W ultraviolet laser on while leaving the redundant shutter closed. A liquid cooling system regulates the temperature of the diode pumps within the laser and requires approximately an hour to stabilize.
10. Check that conditions are safe to open the telescope dome once it is dark enough for observing. This includes a safe range for humidity, dew point depression, precipitation, wind speed, and airborne particles.
11. Open the telescope dome and point to a relatively bright star ($m_V$ ≤ 5) overhead.
12. Refocus the telescope by the positioning the telescope secondary mirror until the star is at approximate best focus (smallest image width). Manual estimation from a live image from one of the science cameras is sufficient.

### 2. High-order Adaptive Optics Correction

1. Pick an astronomical target that has a sufficiently long 'open window' according to the PAM.

2. Set an alarm for the end of the 'open window' with a buffer of at least 1 minute. If the alarm goes off during an observation, immediately shutter the laser.
3. Point the telescope towards the selected astronomical target. Frame the object(s) in the field-of-view of the science cameras by adjusting the telescope pointing as necessary.
4. Confirm that the laser uplink tip-tilt mirror is centered in its range before opening the internal and redundant laser shutters - propagating the laser on sky (**Figure 2**).
5. Record a second of data from the wavefront sensor camera, approximately 1200 frames, while the Pockels cell optical shutter is turned off.
6. Calculate a median image from this data. This will be used as a background frame to subtract any electrical or optical bias from images captured by the wavefront sensor camera.
7. Turn the Pockels cell triggering system on such that the laser pulses from 10 km are transmitted to the wavefront sensor.
8. Spiral search the uplink tip-tilt mirror until the Shack-Hartmann pattern of laser images appear in the wavefront sensor camera (**Figure 4B**). Leave the uplink tip-tilt mirror in position.
9. Record a new wavefront sensor background image while the Pockels cell is momentarily turned off. This is necessary as the optical background changes slightly as the laser is pointed in different directions by the uplink tip-tilt mirror.
10. Start the high-order adaptive-optics system. At this point two control-loops are started simultaneously; the positions of each laser image created by the wavefront sensor lenslet array are used to drive the deformable mirror actuators to flatten the non-planar light waves entering the telescope before they propagate to the science cameras. A weighted average of the position measurements is also used to command the uplink tip-tilt mirror to maintain the centration of the pattern of laser images on the wavefront sensor.

## 3. Observing in the Visible (with Post-facto Registration Correction)

1. Set the position of the filter wheels to the desired observing filter(s).
2. Set the angle of the ADC prisms such that the residual atmospheric prismatic dispersion is minimized on the science instruments.
3. Set the exposure time and frame size on the EMCCD camera such that there is a minimum frame-transfer frame rate of ~10 Hz, with 30 Hz preferred. Data captured at this rate will typically reduce the intra-exposure image motion to below the diffraction-limited angular resolution.
4. Set the electron-multiplication gain on the EMCCD camera such that the maximum intensity of the targets is approximately half the well-depth of the detector or at a maximum value of 300 for the fainter targets.
5. For faint targets, those roughly greater than a stellar magnitude of 15, slow the frame rate of the EMCCD camera down until there are at least ~5-10 photons being detected in the core of the image point spread function. While this leads to additional image motion blurring within frames and reducing angular resolution (*e.g.* ref. [40]; roughly twice the diffraction-limited resolution on $m_r$~16.5 targets), several core photons are required for proper post-facto registration processing.
6. Record a continuous set of images from the EMCCD camera until the total integrated exposure time equals the time calculated in 1.2.

## 4. Observing in the Infrared (with Visible Tip-tilt Correction)

1. Set the filter wheel in front of the EMCCD camera to a broadband filter, *i.e.* a clear filter or a λ > 600nm long-pass filter.
2. Note the pixel position of the object to be used as a tip-tilt guide source on the EMCCD camera while looking at a live image.
3. Set the camera readout settings to the following values: bin pixels by a factor of 4, and set the frame-transfer sub-frame readout region to be a total of 2 × 2 binned pixels centered on the previously noted position.
4. Set the EMCCD camera frame rate and electron multiplication gain to match the brightness of the tip-tilt guide source. A frame rate of 300 Hz is preferred (for a control-loop correction bandwidth of ~30 Hz), but can be lowered as necessary for fainter objects at the cost of lower quality tip-tilt correction.
5. Start the tip-tilt control-loop. This will calculate the current guide source position and command the fast tip-tilt correcting mirror to drive its position to the center of the binned pixel region.
6. Record images from the infrared camera until the total integrated exposure time equals the time calculated in 1.2. Maximum single-frame exposure times will be limited only by saturation from infrared emission, from the sky, instrument or object, or by dark current from the infrared array. Exposures may range from fractions of a second to several minutes.

## 5. End of Night Procedures

1. Close the telescope dome and point the telescope to the flat screen when observing is complete.
2. Turn the laser off and contact USSTRATCOM with a summary of the nightly activities within 15 min.
3. Turn the dome flat lamp on.
4. Record a series of full-frame images on both the EMCCD and infrared cameras of the flat-field illumination produced by the dome flat lamp on the flat screen for each astronomical filter used during the preceding night. The flat-field intensity at each pixel represents the combined relative quantum-efficiency of the telescope, adaptive-optics system, filters and camera.
5. Turn the dome flat lamp off and switch to the blocking filters in front of each camera.
6. Record a series of dark images on both cameras corresponding to the range of exposure times and image formats recorded during the preceding night. The dark frames are used to remove bias due to dark current and electronic noise from recorded data.
7. Park the telescope.

## 6. Processing Images

1. Create a single dark calibration image from the median of each dark image series recorded in 5.6).
2. Create a flat-field calibration image for each filter on each camera by calculating the median of each flat-field image series recorded in 5.4), subtracting the corresponding dark calibration image and then dividing the entire image by the median pixel value in the frame.

3. Subtract the appropriate dark calibration image and divide by the flat-field calibration image for each on-sky science image recorded from the EMCCD and infrared cameras.
4. Re-center the calibrated science images from each observation by aligning the brightest pixel and add the images together to create a stacked image. More sophisticated routines for improved image registration can also be used[39,41].

## Representative Results

The Robo-AO laser adaptive-optics system is used to compensate for atmospheric turbulence and produce diffraction-limited-resolution images at visible and near-infrared wavelengths. **Figure 1A** shows an image of a single star seen in red light through uncompensated atmospheric turbulence with an image width of 1.0 arc second. **Figure 1B** shows the same star after adaptive optics correction: the image width decreases to 0.12 arc seconds, slightly larger than a perfect image width of 0.10 arc seconds at this wavelength on a 1.5-m telescope. The first Airy ring, a result of diffraction, can be seen as the faint ring like structure around the core of the image. This much-improved angular resolution enables the discovery of binary and multiple star systems (*e.g.* **Figure 1C** and observations by ref.[40]) and for the detection of much fainter stars in dense fields such as the globular cluster of Messier 3 (seen in the near-infrared; **Figure 6**) that would otherwise be impossible to directly view through atmospheric turbulence. Features of solar-system objects, such as the cloud surface of Jupiter as well as its transiting moon Ganymede (**Figure 7**), can also be seen with a greater degree of clarity when viewed with laser adaptive optics.

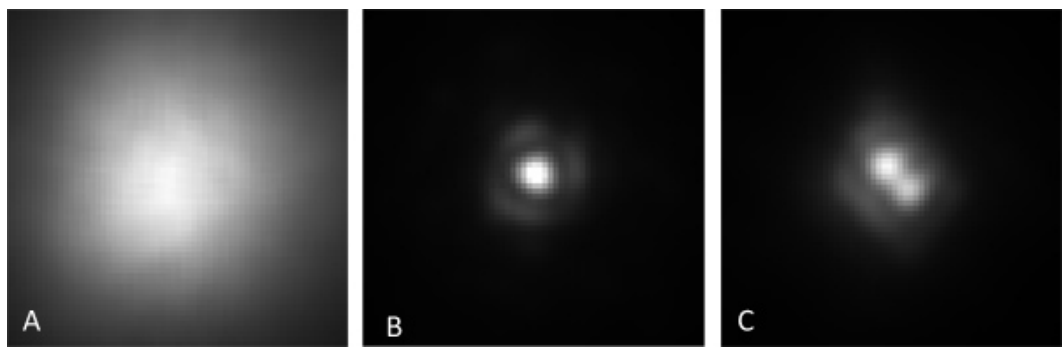

**Figure 1. Adaptive optics correction at visible wavelengths.** Each figure represents a 1.5 × 1.5 arc second field-of-view on sky. **(A)** A long-exposure single image of a single star, $m_V$ =3.5, seen through uncompensated atmospheric turbulence in i-band (λ = 700 - 810 nm) at the 1.5-m P60 telescope at Palomar Observatory. The full-width at half-maximum (FWHM) is 1.0 arc seconds. **(B)** The same star as in **(A)** with laser adaptive optics correction using the Robo-AO system. The core of the stellar image has 15 times the peak brightness of the uncompensated image and has a FWHM of 0.12 arc seconds. **(C)** A binary star, $m_V$ =8.4, with a separation of 0.14 arc seconds is revealed through the use of the Robo-AO adaptive-optics system. In each case, tip-tilt guiding was performed the target itself.

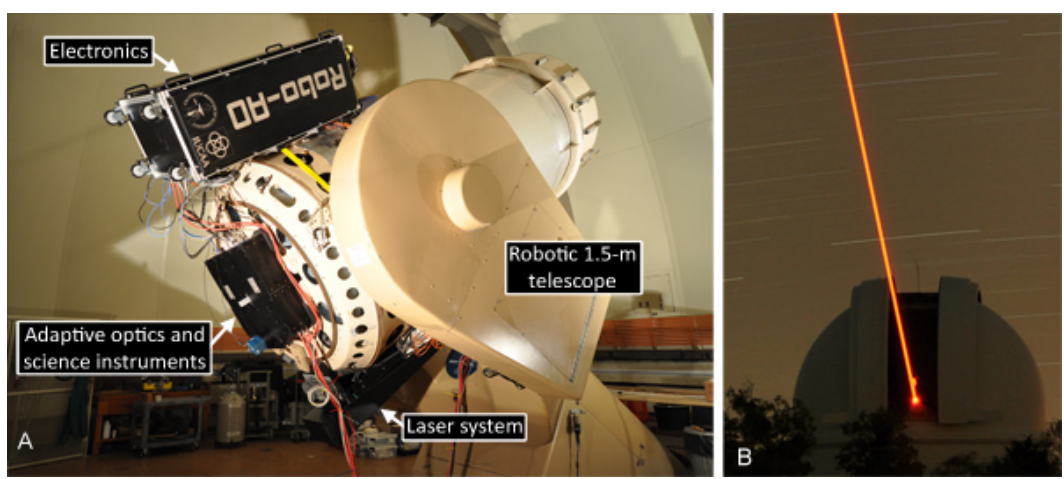

**Figure 2. The Robo-AO laser adaptive-optics system. (A)** The adaptive optics and science instruments are installed at the Cassegrain focus of the robotic 1.5-m P60 telescope at Palomar Observatory. The laser system and support electronics are attached to opposite sides of the telescope tube for balance. **(B)** The Robo-AO UV laser beam propagating out of the telescope dome. In this long exposure photograph, the laser beam is visible due to Rayleigh scattering off of air molecules; a tiny fraction of the light also scatters back toward the telescope to be used as a probe of the atmosphere. The laser beam appears orange because of the way UV light is transmitted through the color filters on the UV sensitive camera used to take the picture. Click here to view larger figure.

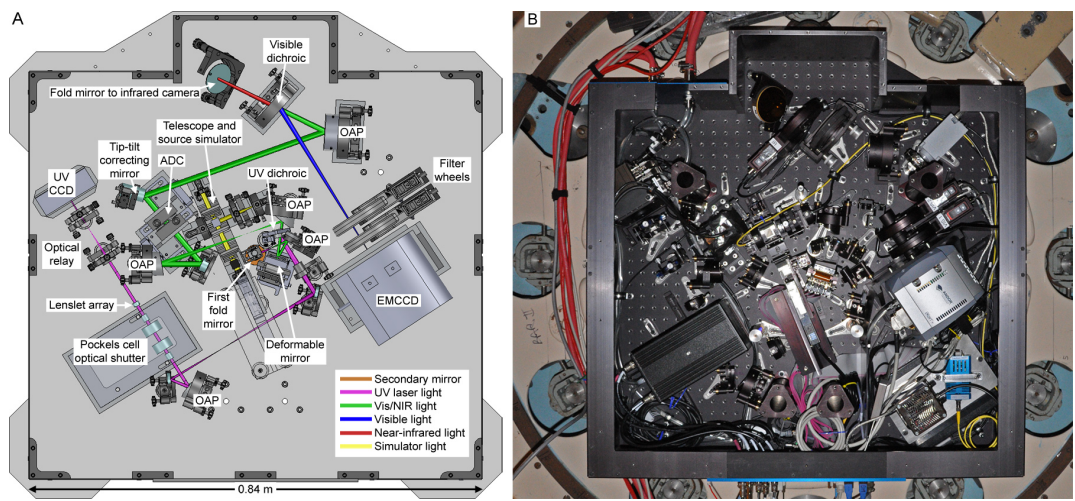

**Figure 3. Robo-AO adaptive optics and science instruments.** (**A**) A simplified CAD model. Light focused from the telescope secondary mirror (*orange*) enters through a small hole at the center of the instrument before being reflected by 90 degrees by the first fold mirror towards an off-axis parabolic (OAP) mirror. This mirror images the telescope pupil on the deformable mirror surface. After reflection from the deformable mirror, an UV dichroic splits off the laser light (*violet*) and directs it to the laser wavefront sensor. An additional reversed OAP mirror within the wavefront sensor corrects the non-common path optical errors introduced by the 10 km conjugate focus of the laser reflecting off of the first OAP mirror. The visible and near-infrared light (*green*) passing through the UV dichroic is relayed by a pair of OAP mirrors to the atmospheric dispersion corrector. The light is then reflected by the tip-tilt correcting mirror to a final OAP mirror which focuses the light towards the visible dichroic. The visible dichroic reflects the visible light (*blue*) to the electron-multiplying CCD and transmits the near-infrared light (*red*) to a fold mirror and ultimately to the infrared camera. The combined UV, visible and near-infrared light from the telescope and source simulator (*yellow*) can be directed to the adaptive optics and science instruments by translating the first fold mirror out of the way. (**B**) A corresponding photograph of the instrument package. Click here to view larger figure.

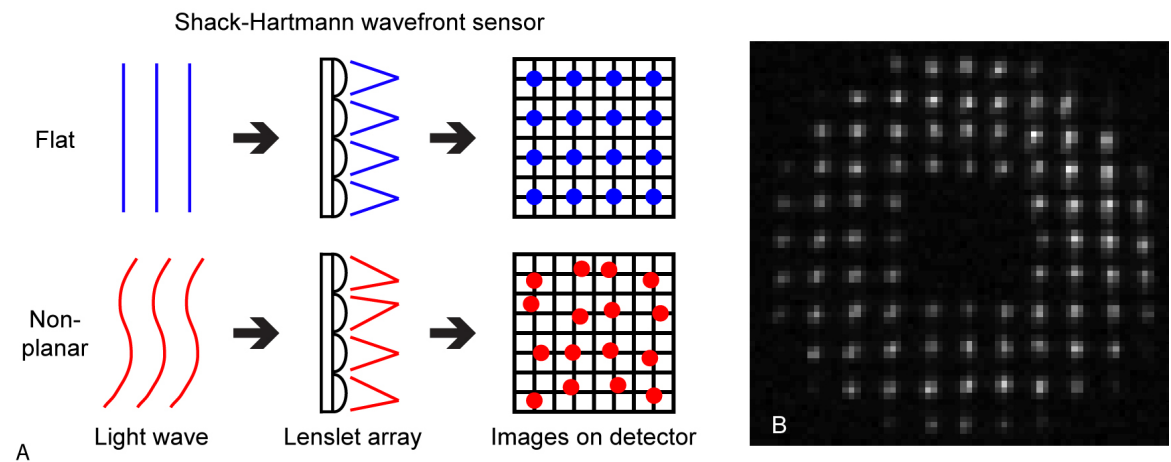

**Figure 4. Shack-Hartmann wavefront sensor.** (**A**) Conceptual diagram. As a flat wave passes through the lenslet array, a regular pattern of images is formed on the detector (*blue*). When a non-planar wave passes through the lenslet array, the local gradient of the wave affects the position of images formed by each lens of the array (*red*). (**B**) Pattern of laser images in the Robo-AO Shack-Hartmann wavefront sensor. Each of the 88 spots is an image of the laser scatter from 10 km as formed by each lens of the lenslet array, with the overall pattern shape determined by the geometry of the telescope pupil. The relative displacement of each image with respect to the reference image position (**Procedure 1.6**) gives a measurement of the local gradient of the incoming light wave. Click here to view larger figure.

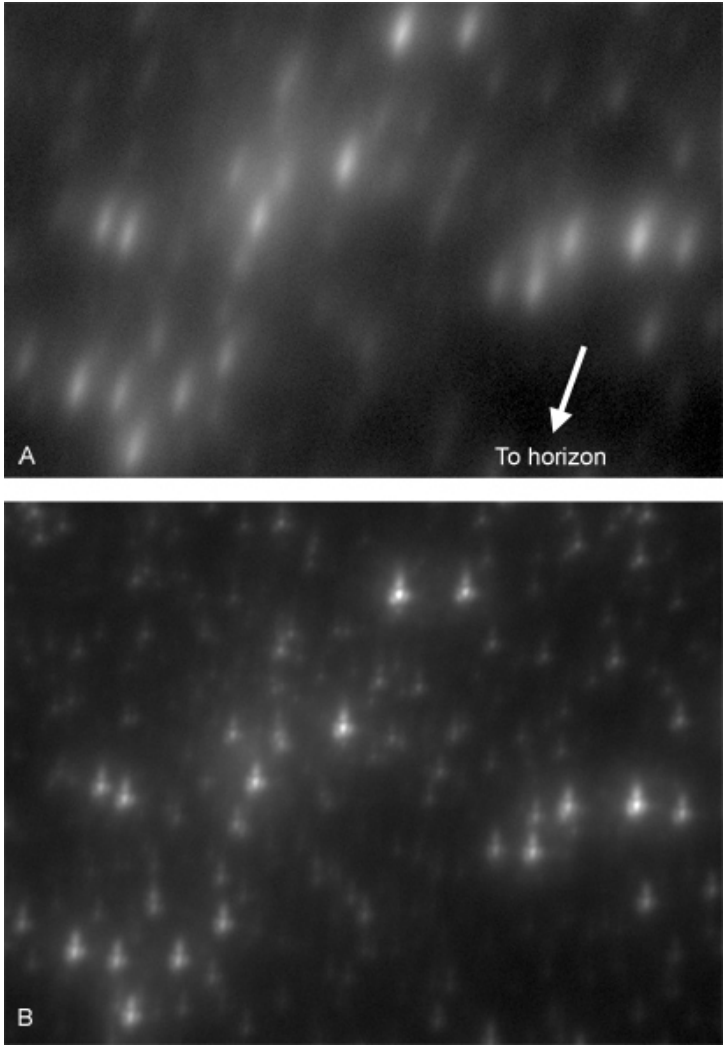

**Figure 5. Correction of atmospheric prismatic dispersion.** Adaptive optics corrected images of an 11 × 16 arc second subfield of the globular cluster Messier 15 at a telescope elevation of 45 degrees. **(A)** While adaptive optics corrects the effects of atmospheric turbulence, atmospheric prismatic dispersion still affects the images of individual stars: images are sharp parallel to the horizon, while elongated perpendicular to the horizon by approximately 1 arc second over a spectral bandwidth of λ = 400 - 950 nm. **(B)** By additionally using an atmospheric dispersion corrector to counteract the atmospheric prismatic dispersion, diffraction-limited-resolution imaging is recovered in both directions.

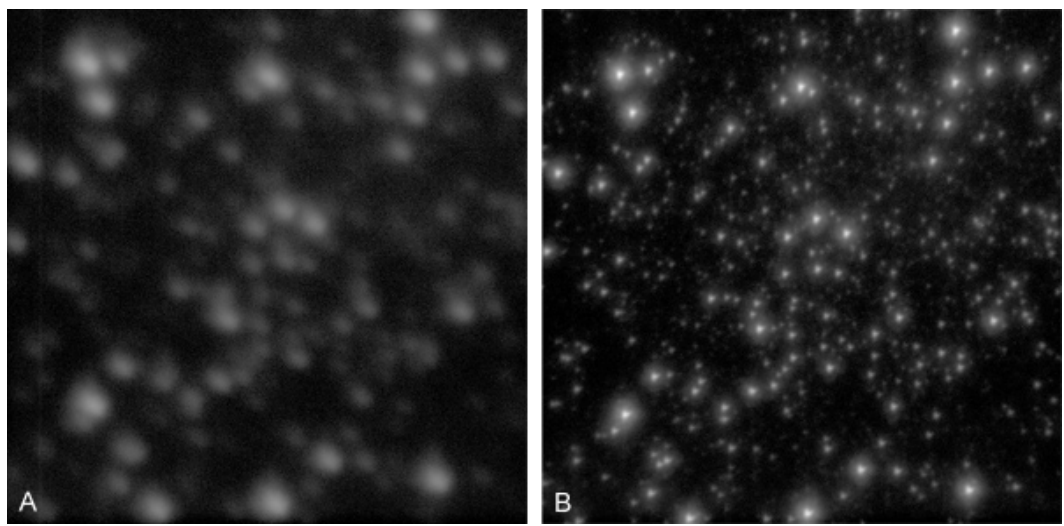

**Figure 6. Images of the globular cluster Messier 3. (A)** A 44 × 44 arc second field-of-view, 2-minute long uncompensated image of the core of the globular cluster Messier 3 in z-band (λ = 830 - 950 nm). **(B)** The same image shown with adaptive optics correction using Robo-AO revealing many stars that could not otherwise be seen.

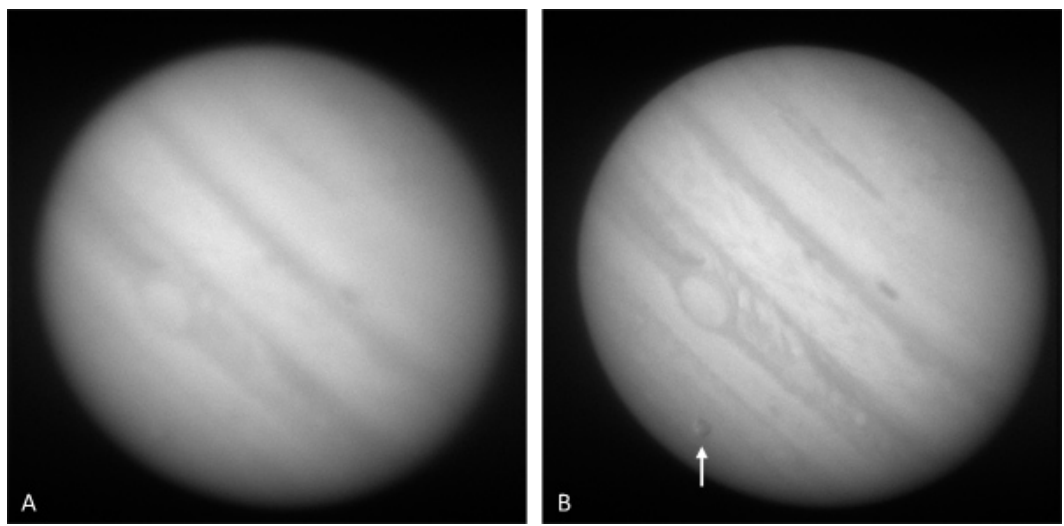

**Figure 7. Images of Jupiter. (A)** A 0.033-second uncompensated snapshot of Jupiter (apparent diameter of 42 arc seconds) in r-band (λ = 560 - 670 nm). **(B)** The same image with Robo-AO laser adaptive optics correction showing the surface cloud features and transiting Ganymede (arrow) with greater clarity.

## Discussion

The method presented here describes the manual operation of the Robo-AO laser adaptive-optics system. In practice, Robo-AO operates in an automated fashion; the vast majority of procedures are controlled by a robotic sequencer which performs the same steps automatically.

The Robo-AO system has been engineered for straightforward replication at modest cost, with materials (~USD600K) and labor being a fraction of the cost of even a 1.5-m telescope. While there are roughly twenty optical telescopes around the world greater than 5 m in diameter, telescopes in the 1-3 m class number well over one-hundred and are projected as potential hosts for Robo-AO clones. In addition to the current system deployed at the 1.5-m P60 telescope, the first of hopefully many clones is being developed for the 2-m IGO telescope[42] in Maharashtra, India, and a variant using bright stars instead of a laser for wavefront sensing is being commissioned at the 1-m telescope at Table Mountain, CA[43]. A revolution in diffraction-limited science may be at hand.

## Disclosures

The authors declare no competing financial interests.

### Acknowledgements

The Robo-AO system is supported by collaborating partner institutions, the California Institute of Technology and the Inter-University Centre for Astronomy and Astrophysics, by the National Science Foundation under Grant Nos. AST-0906060 and AST-0960343, by a grant from the Mt. Cuba Astronomical Foundation and by a gift from Samuel Oschin.

### References

1. Baranec, C. & Dekany, R. Study of a MEMS-based Shack-Hartmann wavefront sensor with adjustable pupil sampling for astronomical adaptive optics. *Applied Optics.* **47**, 5155-5162 (2008).
2. Baranec, C., Riddle, R., Ramaprakash, A.N., Law, N., Tendulkar, S., Kulkarni, S.R., Dekany, R., Bui, K., Davis, J., Burse, M., Das, H., Hildebrandt, S., & Smith, R. Robo-AO: autonomous and replicable laser-adaptive-optics and science system. *Proc. SPIE.* **8447**, 844704 (2012).
3. http://www.astro.caltech.edu/Robo-AO
4. Huygens, C. *The Celestial Worlds discover'd: or, Conjectures concerning the inhabitants, plants and productions of the worlds in the planets.*, (1722).
5. Newton, I. *Opticks.* The Royal Society, (1704).
6. Babcock, H.W. The possibility of compensating astronomical seeing. *Publications of the Astronomical Society of the Pacific.* **65**, 229-236 (1953).
7. Linnik, V.P. On the possibility of reducing atmospheric seeing in the image quality of stars. *Opt. Spectrosc.* **3**, 401-402 (1957).
8. Duffner, R, *The Adaptive Optics Revolution: A History.* Univ. New Mexico Press, Albuquerque, (2009).
9. Fugate, R.Q., ed. Laser Guide Star Adaptive Optics. Proc. Workshop, March 10-12, Starfire Optical Range, Phillips Lab./LITE, Kirtland AFB, NM 87117-6008, (1992).
10. Hardy, J.W. *Adaptive Optics for Astronomical Telescopes.* Oxford, New York, (1998).
11. Hart, M. Recent advances in astronomical adaptive optics. *Applied Optics.* **49**, D17-D29 (2010).
12. Davies, R. & Kasper, M. Adaptive Optics for Astronomy. *Annu. Rev. Astron. Astrophys.* arXiv:astro-ph/1201.5741, In Press (2012).
13. Esposito, S., Riccardi, A., Pinna, E., Puglisi, A., Quirós-Pacheco, F., Arcidiacono, C., Xompero, M., Briguglio, R., Agapito, G., Busoni, L., Fini, L., Argomedo, J., Gherardi, A., Stefanini, P., Salinari, P., Brusa, G., Miller, D., & Guerra, J.C. Large Binocular Telescope Adaptive Optics System: new achievements and perspectives in adaptive optics. *Proc. SPIE.* **8149**, 814902 (2011).
14. Wizinowich, P., Acton, D.S., Shelton, C., Stomski, P., Gathright, J., Ho, K., Lupton, W., Tsubota, K., Lai, O., Max, C., Brase, J., An, J., Avicola, K., Olivier, S., Gavel, D., Macintosh, B., Ghez, A., & Larkin, J. First Light Adaptive Optics Images from the Keck II Telescope: A New Era of High Angular Resolution Imagery. *Publications of the Astronomical Society of the Pacific.* **112**, 315-319 (2000).
15. Minowa, Y., Hayano, Y., Oya, S., Hattori, M., Guyon, O., Egner, S., Saito, Y., Ito, M., Takami, H., Garrel, V., Colley, S., & Golota, T. Performance of Subaru adaptive optics system AO188. *Proc. SPIE.* **7736**, 77363N (2010).
16. Marchetti, E., Brast, R., Delabre, B., Donaldson, R., Fedrigo, E., Frank, C., Hubin, N., Kolb, J., Lizon, J.-L., Marchesi, M., Oberti, S., Reiss, R., Santos, J., Soenke, C., Tordo, S., Baruffolo, A., Bagnara, P., & The CAMCAO Consortium. On-sky Testing of the Multi-Conjugate Adaptive Optics Demonstrator. *The Messenger.* **129**, 8-13 (2007).
17. Rigaut, F., Neichel, B., Bec, M., Boccas, M., d'Orgeville, C., Fesquet, V., Galvez, R., Gausachs, G., Trancho, G., Trujillo, C., Edwards, M., & Carrasco, R. The Gemini Multi-Conjugate Adaptive System sees star light. OSA Conference on Adaptive Optics: Methods, Analysis and Applications, (2011).
18. Hart, M., Milton, N.M., Baranec, C., Powell, K., Stalcup, T., McCarthy, D., Kulesa, C., & Bendek, E. A ground-layer adaptive optics system with multiple laser guide stars. *Nature.* **466**, 727-729 (2010).
19. Troy, M., Dekany, R.G., Brack, G., Oppenheimer, B.R., Bloemhof, E.E., Trinh, T., Dekens, F.G., Shi, F., Hayward, T.L., & Brandl, B. Palomar adaptive optics project: status and performance. *Proc. SPIE.* **Vol. 4007**, 31-40 (2000).
20. Cenko, S.B., Fox, D.B., Moon, D.-S., Harrison, F.A., Kulkarni, S.R., Henning, J.R., Guzman, C.D., Bonati, M., Smith, R.M., Thicksten, R.P., Doyle, M.W., Petrie, H.L., Gal-Yam, A., Soderberg, A.M., Anagnostou, N.L., & Laity, A.C. The automated Palomar 60 inch telescope. *Publications of the Astronomical Society of the Pacific.* **118**, 1396-1406 (2006).
21. Shporer, A., Brown, T., Lister, T., Street, R., Tsapras, Y., Bianco, F., Fulton, B., & Howell, A. The LCOGT Network. The Astrophysics of Planetary Systems: Formation, Structure, and Dynamical Evolution. *Proceedings of the International Astronomical Union, IAU Symposium.* **276**, 553-555 (2011).
22. Baranec, C., Dekany, R., Kulkarni, S., Law, N., Ofek, E., Kasliwal, M., & Velur, V. Deployment of low-cost replicable laser adaptive optics on 1-3 meter class telescopes. *Astro2010: The Astronomy and Astrophysics Decadal Survey.*, (2009).
23. Ofek, E.O., Law, N., & Kulkarni, S.R. Mass makeup of galaxies. *Astro2010: The Astronomy and Astrophysics Decadal Survey.*, (2009).
24. Morton, T.D. & Johnson, J.A. On the low false positive probabilities of Kepler planet candidates. *The Astrophysical Journal.* **738**, 170 (2011).
25. Erickcek, A.L. & Law, N.M. Astrometric microlensing by local dark matter subhalos. *The Astrophysical Journal.* **729**, 49 (2011).
26. Law, N.M., Kulkarni, S.R., Dekany, R.G., & Baranec, C. Planets around M-dwarfs - astrometric detection and orbit characterization. *Astro2010: The Astronomy and Astrophysics Decadal Survey.*, (2009).
27. Kulkarni, S.R. Cosmic Explosions (Optical Transients)., arXiv:1202.2381, (2012).
28. Goldstein, R. Pockels Cell Primer. Laser Focus, Feb., 21-26 (1968).
29. Zigman, S. Effects of near ultraviolet radiation on the lens and retina. *Documenta Ophthalmologica.* **55**, 375-391 (1983).
30. *ANSI Z136.1-2007, American National Standard for Safe Use of Lasers.*, Laser Institute of America, Orlando, (2007).
31. Thompson, L.A. & Teare, S.W. Rayleigh laser guide star systems: application to the University of Illinois seeing improvement system. *Publications of the Astronomical Society of the Pacific.* **114**, 1029-1042 (2002).
32. Department of Defense Instruction 3100.11. Illumination of objects in space by lasers. March 31, (2000).
33. Platt, B.C. & Shack, R. History and principles of Shack-Hartmann wavefront sensing. *J. Refract. Surg.* **17**, S573-S577 (2001).
34. Milton, N.M. & Lloyd-Hart, M., Disk Harmonic Functions for Adaptive Optics Simulations. OSA conference on Adaptive Optics: Analysis and Methods, (2005).

35. Bifano, T., Cornelissen, S., & Bierden, P. MEMS deformable mirrors in astronomical adaptive optics. 1st AO4ELT conference, 06003, doi:10.1051/ao4elt/201006003, (2010).
36. Rigaut, F. & Gendron, G. Laser guide star in adaptive optics: the tilt determination problem. *Astron. Astrophys.* **261**, 677-684 (1992).
37. Devaney, N., Goncharov, A.V., & Dainty, J.C. Chromatic effects of the atmosphere on astronomical adaptive optics. *Applied Optics.* **47**, 8 (2008).
38. Basden, A.G., Haniff, C.A., & Mackay, C.D. Photon counting strategies with low-light-level CCDs. *Monthly Notices of the Royal Astronomical Society.* **345**, 985-991 (2003).
39. Law, N.M., Mackay, C.D., & Baldwin, J.E. Lucky imaging: high angular resolution imaging in the visible from the ground. *Astronomy and Astrophysics.* **446**, 739-745 (2006).
40. Law, N.M., Kraus, A.L., Street, R., Fulton, B.J., Hillenbrand, L.A., Shporer, A., Lister, T., Baranec, C., Bloom, J.S., Bui, K., Burse, M.P., Cenko, S.B., Das, H.K., Davis, J.C.T., Dekany, R.G., Filippenko, A.V., Kasliwal, M.M., Kulkarni, S.R., Nugent, P., Ofek, E.O., Poznanski, D., Quimby, R.M., Ramaprakash, A.N., Riddle, R., Silverman, J.M., Sivanandam, S., & Tendulkar, S. Three new eclipsing white-dwarf - M-dwarf binaries discovered in a search for transiting Planets around M-dwarfs. *The Astrophysical Journal.* arXiv:1112.1701, In Press, (2012).
41. Law, N.M., Mackay, C.D., Dekany, R.G., Ireland, M., Lloyd, J.P., Moore, A.M., Robertson, J.G., Tuthill, P., & Woodruff, H.C. Getting Lucky with Adaptive Optics: Fast Adaptive Optics Image Selection in the Visible with a Large Telescope. *The Astrophysical Journal.* **692**, 924-930 (2009).
42. Gupta, R., Burse, M., Das, H.K., Kohok, A., Ramaprakash, A.N., Engineer, S., & Tandon, S.N. IUCAA 2 meter telescope and its first light instrument IFOSC. *Bulletin of the Astronomical Society of India*. **30**, 785 (2002).
43. Choi, P.I., Severson, S.A., Rudy, A.R., Gilbreth, B.N., Contreras, D.S., McGonigle, L.P., Chin, R.M., Horn, B., Hoidn, O., Spjut, E., Baranec, C., & Riddle, R. KAPAO: a Pomona College adaptive optics instrument. *Bulletin of the American Astronomical Society.* **43**, (2011).